\documentclass[letterpaper, 10 pt, conference]{ieeeconf}%
\usepackage{bm}
\usepackage{amssymb}
\usepackage{epsfig}
\usepackage{subfigure}
\usepackage{amsmath}
\usepackage{xcolor}
\usepackage{graphicx}
\usepackage{epstopdf}
\usepackage{amsfonts}
\usepackage{indentfirst}%
\usepackage[ruled]{algorithm2e}
\providecommand{\U}[1]{\protect\rule{.1in}{.1in}}
\newtheorem{problem}{\textbf{Problem}}
\newtheorem{definition}{\textbf{Definition}}

\newtheorem{theorem}{\rm\textbf{Theorem}}

\newtheorem{assump}{\rm\textbf{Assumption}}
\newtheorem{remark}{\rm\textbf{Remark}}
\IEEEoverridecommandlockouts
\begin{document}

\title{{\LARGE \textbf{Event-Triggered Safety-Critical Control for Systems\\ with Unknown Dynamics}}}
\author{Wei Xiao, Calin Belta and Christos G. Cassandras\thanks{This work was
supported in part by NSF under grants ECCS-1931600, DMS-1664644, CNS-1645681, IIS-1723995, and IIS-2024606, by ARPAE under grant DE-AR0001282 and by its NEXTCAR program under grant DE-AR0000796, by AFOSR under grant FA9550-19-1-0158, and by the MathWorks. The
authors are with the Division of Systems Engineering and Center for
Information and Systems Engineering, Boston University, Brookline, MA, 02446,
USA \texttt{{\small \{xiaowei\}@bu.edu}}}}
\maketitle

\begin{abstract}
This paper addresses the problem of safety-critical control for systems with unknown dynamics. It has been shown that stabilizing affine control systems to desired (sets of)
states while optimizing quadratic costs subject to state and control
constraints can be reduced to a sequence of quadratic programs (QPs) by using
Control Barrier Functions (CBFs) and Control Lyapunov Functions (CLFs). Our recently proposed High Order CBFs (HOCBFs) can accommodate constraints of arbitrary relative degree. One of the main challenges in this approach is obtaining accurate system dynamics, which is especially difficult for systems that require online model identification given limited computational resources and system data.  
In order to approximate the real unmodelled system dynamics, 
we define adaptive affine control dynamics which are updated based on the error states obtained by real-time sensor measurements. 
We define a HOCBF for a safety requirement on the unmodelled system based on the adaptive dynamics and error states, and reformulate the safety-critical control problem as the above mentioned QP. Then, we determine the events required to solve the QP in order to guarantee safety. We also derive a condition that guarantees the satisfaction of the HOCBF constraint between events.  We illustrate the effectiveness of the proposed framework on an adaptive cruise control problem and compare it with the classical time-driven approach.
\end{abstract}

\thispagestyle{empty} \pagestyle{empty}



\section{INTRODUCTION}

\label{sec:intro}

Constrained optimal control problems with safety specifications are central to increasingly widespread safety critical autonomous and cyber physical systems. Traditional Hamiltonian analysis \cite{Bryson1969} and dynamic programming \cite{Denardo2003} cannot accommodate the size and nonlinearities of such systems, and work efficiently for small-scale linear systems. 
Model predictive control (MPC) \cite{Rawlings2018} methods have been shown to work for large, non-linear systems that can be easily linearized. However, safety requirements are hard to guarantee. Motivated by these limitations, barrier and control barrier functions enforcing safety have received increased attention in recent years \cite{Aaron2014} \cite{Glotfelter2017} \cite{Xiao2019}. 

Barrier functions (BFs) are Lyapunov-like functions \cite{Tee2009},
\cite{Wieland2007}, whose use can be traced back to optimization problems
\cite{Boyd2004}. More recently, they have been employed to prove set
invariance \cite{Aubin2009}, \cite{Prajna2007}, \cite{Wisniewski2013} and for
multi-objective control \cite{Panagou2013}. In \cite{Tee2009}, it was proved
that if a BF for a given set satisfies Lyapunov-like conditions, then the set
is forward invariant. A less restrictive form of a BF, which is allowed to
grow when far away from the boundary of the set, was proposed in
\cite{Aaron2014}. Another approach that allows a BF to be zero was proposed in
\cite{Glotfelter2017}, \cite{Lindemann2018}. This simpler form has also been
considered in time-varying cases and applied to enforce Signal Temporal Logic
(STL) formulas as hard constraints \cite{Lindemann2018}.

Control BFs (CBFs) are extensions of BFs for control systems, and are used to
map a constraint defined over system states to a constraint on the control
input. The CBFs from \cite{Aaron2014} and \cite{Glotfelter2017} work for constraints
that have relative degree one with respect to the system dynamics. A
backstepping approach was introduced in \cite{Hsu2015} to address higher
relative degree constraints, and it was shown to work for relative degree two.
A CBF method for position-based constraints with relative degree two was also
proposed in \cite{Wu2015}. A more general form \cite{Nguyen2016}, which works
for arbitrarily high relative degree constraints, employs input-output
linearization and finds a pole placement controller with negative poles to
stabilize an exponential CBF to zero. The high order
CBF (HOCBF) proposed in \cite{Xiao2019} is simpler and more general than the
exponential CBF \cite{Nguyen2016}.

Most works using CBFs to enforce safety are based on the assumption that the control system is affine in controls and the cost is quadratic in controls. Convergence to desired states is achieved by Control Lyapunov Functions (CLFs) \cite{Aaron2012}.  The time domain is discretized, and the state is assumed to be constant over each time interval. 
The optimal control problem becomes a Quadratic Program (QP) in each time interval and the control is kept constant for the whole interval. Using this approach, the original optimal control problem is reduced to a (possibly large) sequence of quadratic programs (QP) - one for each interval \cite{Galloway2013}. One of the challenges in this QP-based approach is to determine the next time  to solve the QP such that safety can still be guaranteed due to time discretization. The work in \cite{Yang2019} proposed to find the next time to solve the QP by considering the system Lipschitz constants, and the work in \cite{Taylor2021} used a similar idea as the event-triggered control for Lyapunov functions \cite{Tabuada2007}. All these approaches assume that the dynamics are accurately modelled, which is often not the case in reality.   

In order to find accurate dynamics for systems with uncertainties, \cite{Taylor2020} proposed to use machine learning techniques; this, however, is computationally expensive and is not guaranteed to yield sufficiently accurate dynamics for the CBF method. The work in \cite{Sadra2018} proposed to use piecewise linear systems to estimate the system dynamics, which is also computationally expensive. All these works fail to work for systems (such as time-varying systems) that require online model identification.

In order to address the problem of safety-critical control for systems with unknown dynamics, especially for systems that are hard to be accurately modelled and require online identification, the contribution of this paper is to define adaptive affine dynamics that are updated in a time-efficient way to approximate the actual unmodelled dynamics. The adaptive and real dynamics are related through the error states obtained by real-time sensor measurements. We define a HOCBF for a safety requirement on the actual system based on the adaptive dynamics and error states, and reformulate the safety-critical control problem as the above mentioned QP. We determine the events required to solve the QP in order to guarantee safety and derive a condition that guarantees the satisfaction of the HOCBF constraint between events. The adaptive dynamics are updated at each event to accommodate the real dynamics according to the error states; this can reduce the number of events, thus improving the computational efficiency. Our framework can accommodate measurement uncertainties, guarantee safety for systems with unknown dynamics, and is more time efficient than the approaches in \cite{Taylor2020}, \cite{Sadra2018}.  We illustrate our approach and compare with the classical time driven method on an ACC problem.


\section{PRELIMINARIES}

\label{sec:pre}

\begin{definition}
\label{def:classk} (\textit{Class $\mathcal{K}$ function} \cite{Khalil2002}) A
continuous function $\alpha:[0,a)\rightarrow[0,\infty), a > 0$ is said to
belong to class $\mathcal{K}$ if it is strictly increasing and $\alpha(0)=0$.
\end{definition}

Consider an affine control system (assumed to be known in this section) of the form 
\begin{equation}
\dot{\bm{x}}=f(\bm x)+g(\bm x)\bm u \label{eqn:affine}%
\end{equation}
where $\bm x\in X\subset\mathbb{R}^{n}$, $f:\mathbb{R}^{n}\rightarrow\mathbb{R}^{n}$
and $g:\mathbb{R}^{n}\rightarrow\mathbb{R}^{n\times q}$ are
Lipschitz continuous, and $\bm u\in U\subset\mathbb{R}^{q}$ is the control constraint set
defined as
\begin{equation}
U:=\{\bm u\in\mathbb{R}^{q}:\bm u_{min}\leq\bm u\leq\bm u_{max}\}.
\label{eqn:control}%
\end{equation}
with $\bm u_{min},\bm u_{max}\in\mathbb{R}^{q}$ and the inequalities are
interpreted componentwise.

\begin{definition}
\label{def:forwardinv} A set $C\subset\mathbb{R}^{n}$ is forward invariant for
system (\ref{eqn:affine}) if its solutions starting at any $\bm x(0) \in C$
satisfy $\bm x(t)\in C,$ $\forall t\geq0$.
\end{definition}

\begin{definition}
\label{def:relative} (\textit{Relative degree}) The relative degree of a
(sufficiently many times) differentiable function $b:\mathbb{R}^{n}%
\rightarrow\mathbb{R}$ with respect to system (\ref{eqn:affine}) is the number
of times it needs to be differentiated along its dynamics until the control
$\bm u$ explicitly shows in the corresponding derivative.
\end{definition}

In this paper, since function $b$ is used to define a constraint $b(\bm
x)\geq0$, we will also refer to the relative degree of $b$ as the relative
degree of the constraint. 

For a constraint $b(\bm x)\geq0$ with relative
degree $m$, $b:\mathbb{R}^{n}\rightarrow\mathbb{R}$, and $\psi_{0}(\bm
x):=b(\bm x)$, we define a sequence of functions $\psi_{i}:\mathbb{R}%
^{n}\rightarrow\mathbb{R},i\in\{1,\dots,m\}$:
\begin{equation}
\begin{aligned} \psi_i(\bm x) := \dot \psi_{i-1}(\bm x) + \alpha_i(\psi_{i-1}(\bm x)),i\in\{1,\dots,m\}, \end{aligned} \label{eqn:functions}%
\end{equation}
where $\alpha_{i}(\cdot),i\in\{1,\dots,m\}$ denotes a $(m-i)^{th}$ order
differentiable class $\mathcal{K}$ function.

We further define a sequence of sets $C_{i}, i\in\{1,\dots,m\}$ associated
with (\ref{eqn:functions}) in the form:
\begin{equation}
\label{eqn:sets}\begin{aligned} C_i := \{\bm x \in \mathbb{R}^n: \psi_{i-1}(\bm x) \geq 0\}, i\in\{1,\dots,m\}. \end{aligned}
\end{equation}

\begin{definition}
\label{def:hocbf} (\textit{High Order Control Barrier Function (HOCBF)}
\cite{Xiao2019}) Let $C_{1}, \dots, C_{m}$ be defined by (\ref{eqn:sets}%
) and $\psi_{1}(\bm x), \dots, \psi_{m}(\bm x)$ be defined by
(\ref{eqn:functions}). A function $b: \mathbb{R}^{n}\rightarrow\mathbb{R}$ is
a High Order Control Barrier Function (HOCBF) of relative degree $m$ for
system (\ref{eqn:affine}) if there exist $(m-i)^{th}$ order differentiable
class $\mathcal{K}$ functions $\alpha_{i},i\in\{1,\dots,m-1\}$ and a class
$\mathcal{K}$ function $\alpha_{m}$ such that 
{\small\begin{equation}
\label{eqn:constraint}\begin{aligned} \sup_{\bm u\in U}[L_f^{m}b(\bm x) + L_gL_f^{m-1}b(\bm x)\bm u + R(b(\bm x)) + \alpha_m(\psi_{m-1}(\bm x))] \geq 0, \end{aligned}
\end{equation}
}for all $\bm x\in C_{1} \cap,\dots, \cap C_{m}$. In
(\ref{eqn:constraint}), $L_{f}^{m}$ ($L_{g}$) denotes Lie derivatives along
$f$ ($g$) $m$ (one) times, and $R(\cdot)$ denotes the remaining Lie derivatives
along $f$ with degree less than or equal to $m-1$ (omitted for simplicity, see
\cite{Xiao2019}).
\end{definition}

The HOCBF is a general form of the relative degree one CBF \cite{Aaron2014},
\cite{Glotfelter2017}, \cite{Lindemann2018}, i.e., setting $m=1$ reduces the HOCBF to
the common CBF form:
\begin{equation}\label{eqn:cbf0}
L_fb(\bm x) + L_gb(\bm x)\bm u + \alpha_1(b(\bm x))\geq 0,
\end{equation}
 and it is also a general form of the exponential CBF
\cite{Nguyen2016}.

\begin{theorem}
\label{thm:hocbf} (\cite{Xiao2019}) Given a HOCBF $b(\bm x)$ from Def.
\ref{def:hocbf} with the associated sets $C_{1}, \dots, C_{m}$ defined
by (\ref{eqn:sets}), if $\bm x(0) \in C_{1} \cap,\dots,\cap C_{m}$,
then any Lipschitz continuous controller $\bm u(t)$ that satisfies
(\ref{eqn:constraint}), $\forall t\geq0$ renders $C_{1}\cap,\dots,
\cap C_{m}$ forward invariant for system (\ref{eqn:affine}).
\end{theorem}

\begin{definition}
\label{def:clf} (\textit{Control Lyapunov Function (CLF)} \cite{Aaron2012}) A
continuously differentiable function $V: \mathbb{R}^{n}\rightarrow\mathbb{R}$
is an exponentially stabilizing control Lyapunov function (CLF) for system
(\ref{eqn:affine}) if there exist constants $c_{1} >0, c_{2}>0, c_{3}>0$ such
that for $\forall\bm x\in\mathbb{R}^{n}$, $c_{1}||\bm x||^{2} \leq V(\bm x)
\leq c_{2} ||\bm x||^{2}, $
\begin{equation} \label{eqn:clf}
\inf_{\bm u\in U} \lbrack L_{f}V(\bm x)+L_{g}V(\bm x)
\bm u + c_{3}V(\bm x)\rbrack\leq0.
\end{equation}

\end{definition}

Many existing works \cite{Aaron2014}, \cite{Nguyen2016}, \cite{Yang2019}
combine CBFs for systems with relative degree one with quadratic costs to form
optimization problems. An explicit solution to such problems can be obtained based on some assumptions \cite{Ames2017}. Alternatively, we can discretize time and an optimization problem with
constraints given by the CBFs (inequalities of the form (\ref{eqn:constraint}%
)) is solved at each time step. The inter-sampling effect in this approach is considered in \cite{Yang2019}. If convergence to a state is desired, then a
CLF constraint of the form (\ref{eqn:clf}) is added, as in \cite{Aaron2014} \cite{Yang2019}. Note that these
constraints are linear in control since the state value is fixed at the
beginning of the interval. Therefore, each optimization problem is a quadratic
program (QP). The optimal control obtained by solving each QP is applied at
the current time step and held constant for the whole interval. The state is
updated using dynamics (\ref{eqn:affine}), and the procedure is repeated. 
Replacing CBFs by HOCBFs allows us to handle constraints with arbitrary
relative degree \cite{Xiao2019}. Throughout the paper, we will refer to this method as the {\it time driven} approach. The CBF method works if (\ref{eqn:affine}) is an accurate model for the system. However, this is often not the case in reality, especially for time-varying systems. In what follows, we show
how we can find a safety-guaranteed controller for systems with unknown dynamics.

\section{PROBLEM FORMULATION AND APPROACH}
\label{sec:prob}
\begin{figure}[thpb]
	\centering
	\includegraphics[scale=0.3]{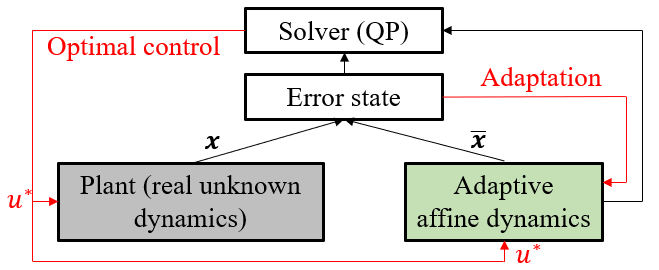}
	\caption{The solution framework for Problem \ref{prob:general} and the connection between the real unknown dynamics and the adaptive affine dynamics (\ref{eqn:affine_nom}). The state $\bm x$ is from the sensor measurements of the plant.}
	\label{fig:fk}	
\end{figure}

We consider a system (state $\bm x \in\mathbb{R}^n$ and control $\bm u \in U$) with unknown dynamics, as shown in Fig. \ref{fig:fk}. For the unknown dynamics, we make the following assumption:
\begin{assump}\label{asp:rd}
	The relative degree of each component of $\bm x$ is known with repect to the real unknown dynamics\footnote{The relative degree is defined similarly to Def. \ref{def:relative} for the real unknown dynamics.}.
\end{assump}

For example, if the position of a vehicle (whose dynamics are unknown) is a component in $\bm x$ and the control is acceleration, then the relative degree of the position with respect to the unknown vehicle dynamics is two by Newton's law. We assume that we have sensors to monitor $\bm x$ and its derivatives.

$\textbf{Objective}$: (Minimizing cost) Consider an optimal control problem for the real unknown dynamics with the cost:
\begin{equation}\label{eqn:cost}
\min_{\bm u(t)}\int_{0}^{T}\mathcal{C}(||\bm u(t)||)dt + p_0||\bm x(T) - \bm K||^2
\end{equation}
where $||\cdot||$ denotes the 2-norm of a vector, $\mathcal{C}(\cdot)$ is a
strictly increasing function of its argument. $T > 0, p_0 > 0, \bm K\in\mathbb{R}^n$.

$\textbf{Safety requirements}$: 
The real unknown dynamics should always satisfy a safety requirement:
\begin{equation} \label{eqn:safetycons}
b(\bm x(t))\geq 0, \forall t\in[0,T].
\end{equation}
where $b: \mathbb{R}^n\rightarrow\mathbb{R}$ is continuously differentiable and has relative degree $m\in\mathbb{N}$ with respect to the real system. The relative degree $m$ is known by Assumption \ref{asp:rd}.

$\textbf{Control constraints}$: The control of the real system should always satisfy control bounds in the form of (\ref{eqn:control}).

A control policy for the real unknown dynamics is $\bm {feasible}$ if constraints (\ref{eqn:safetycons}) and (\ref{eqn:control}) are satisfied for all times. Note that state limitations are particular forms of (\ref{eqn:safetycons}). In this paper, we consider the following problem:

\vspace{2mm}
\begin{problem}\label{prob:general}
	Find a feasible control policy for the real unknown dynamics such that the cost (\ref{eqn:cost}) is minimized. 
\end{problem}
\vspace{2mm}

\textbf{Approach:} Our approach to solve Problem \ref{prob:general} relies on the CBF-based QP method \cite{Aaron2014}. There are four steps involved in the solution:

\textbf{Step 1: define adaptive affine dynamics.} Our motivation is that we need affine dynamics of the form (\ref{eqn:affine}) in order to apply the CBF-based QP approach to solve Problem \ref{prob:general}. Under Assump. \ref{asp:rd}, we define affine dynamics that have the same relative degree for (\ref{eqn:safetycons}) as the real system to estimate the real unknown dynamics in the form: 
\begin{equation}
\dot{\bar{\bm{x}}}=f_a(\bar{\bm{x}})+g_a(\bar{\bm{x}})\bm u \label{eqn:affine_nom}%
\end{equation}
where $f_a:\mathbb{R}^n\rightarrow \mathbb{R}, g_a:\mathbb{R}^{n}\rightarrow\mathbb{R}^{n\times q}$, and $\bar{\bm{x}}\in X\subset \mathbb{R}^n$ is the state vector corresponding to $\bm x$ in the unknown dynamics. Since $f_a(\cdot), g_a(\cdot)$ in (\ref{eqn:affine_nom}) can be adaptively updated to accommodate the real unknown dynamics, as shown in the next section, we call (\ref{eqn:affine_nom}) {\it adaptive affine dynamics}. The real unknown dynamics and (\ref{eqn:affine_nom}) are related through the error states obtained from the real-time measurements of the system and the integration of (\ref{eqn:affine_nom}). Theoretically, we can take any affine dynamics in (\ref{eqn:affine_nom}) to model the real system as long as their states are of the same dimension and with the same physical interpretation within the plant. Clearly, we would like the adaptive dynamics (\ref{eqn:affine_nom}) to ``stay close'' to the real dynamics. This notion will be formalized in the next section.

\textbf{Step 2: find a HOCBF that guarantees (\ref{eqn:safetycons}).}
Based on (\ref{eqn:affine_nom}), the error state and its derivatives, we use a HOCBF to enforce (\ref{eqn:safetycons}). Details are shown in the next section. 

\textbf{Step 3: formulate the CBF-based QP.}
We use a relaxed CLF to achieve a minimal value of the terminal state penalty in (\ref{eqn:cost}). If $\mathcal{C}(||\bm u(t)||) =||\bm u(t)||^2 $ in (\ref{eqn:cost}), then we can formulate Problem \ref{prob:general} using a CBF-CLF-QP approach \cite{Aaron2014}, 
with a CBF replaced by a HOCBF \cite{Xiao2019} if $m > 1$. 

\textbf{Step 4: determine the events required to solve the QP and the condition that guarantees the satisfaction of (\ref{eqn:safetycons}) between events.} Since there is obviously a difference between the adaptive affine dynamics (\ref{eqn:affine_nom}) and the real unknown dynamics, in order to guarantee safety in the real system, we need to properly define events (dependent on the error state and the state of (\ref{eqn:affine_nom})) to solve the QP. In other words, we need to determine the times $t_k, k = 1, 2, \dots (t_1 = 0)$ at which the QP  must be solved in order to guarantee the satisfaction of (\ref{eqn:safetycons}) for the real unknown dynamics.

 The proposed solution framework is shown in Fig. \ref{fig:fk} where we note that we apply the same control from the QP to both the real unknown dynamics and (\ref{eqn:affine_nom}).  
 
\section{Event-Triggered Control}
\label{sec:etc}

In this section, we provide the technical details involved in formulating the CBF-based QPs that guarantee the satisfaction of the safety constraint (\ref{eqn:safetycons}) for the real unknown system.
We start with the case of a relative-degree-one safety constraint (\ref{eqn:safetycons}).

\subsection{Relative-degree-one Constraints}

Suppose the safety constraint in (\ref{eqn:safetycons}) has relative degree one with respect to both dynamics (\ref{eqn:affine_nom}) and the actual dynamics. 

Next, we show how to find a CBF that guarantees (\ref{eqn:safetycons}) for the real unknown dynamics.
Let
\begin{equation} \label{eqn:err}
\bm e := \bm x - \bar{\bm x}.
\end{equation} 
Note that $\bm x$ and $\bar{\bm x}$ are state vectors from direct measurements and from the adaptive dynamics (\ref{eqn:affine_nom}), respectively. Then, 
\begin{equation}\label{eqn:alt}
b(\bm x) = b(\bar{\bm x} + \bm e).
\end{equation} 

Differentiating $b(\bar{\bm x} + \bm e)$, we have
\begin{equation} \label{eqn:deri}
\begin{aligned}
 \frac{db(\bar{\bm x} + \bm e)}{dt} 
 =\frac{\partial b(\bar{\bm x} + \bm e)}{\partial \bar{\bm x}} \dot{\bar{\bm x}}  + \frac{\partial b(\bar{\bm x} + \bm e)}{\partial \bm e} \dot{\bm e}
\end{aligned}
\end{equation}

The CBF constraint  that guarantees (\ref{eqn:safetycons}) for known dynamics (\ref{eqn:affine}) is as in (\ref{eqn:cbf0}), which is done by replacing $\dot {\bm x}$ with (\ref{eqn:affine}). However, for the unknown dynamics, the CBF constraint is:
$
\frac{db(\bm x)}{dt} + \alpha_1(b(\bm x))\geq 0.
$
Equivalently, we have 
\begin{equation} \label{eqn:cbf}
\frac{db(\bar{\bm x} + \bm e)}{dt} + \alpha_1(b(\bar{\bm x} + \bm e))\geq 0.
\end{equation}

Combining (\ref{eqn:deri}), (\ref{eqn:cbf}) and (\ref{eqn:affine_nom}), we get the CBF constraint that guarantees (\ref{eqn:safetycons}):
\begin{equation} \label{eqn:cbf_re}
\frac{\partial b(\bm x)}{\partial\bar{\bm x}} f_a(\bar{\bm x}) + \frac{\partial b(\bm x)}{\partial\bar{\bm x}} g_a(\bar{\bm x})\bm u + \frac{\partial b(\bm x)}{\partial\bm e} \dot{\bm e} + \alpha_1(b(\bm x))\geq 0.
\end{equation}
where $\dot {\bm e} = \dot{\bm x} - \dot{\bar{\bm x}}$ is evaluated online through $\dot {\bm x}$ (from direct measurements of the actual state derivative) and $\dot{\bar{\bm x}}$ is given through (\ref{eqn:affine_nom}). Then, the satisfaction of (\ref{eqn:cbf_re}) implies the satisfaction of $b(\bar{\bm x} + \bm e) \geq 0$ by Thm. \ref{thm:hocbf} and (\ref{eqn:alt}), therefore, (\ref{eqn:safetycons}) is guaranteed to be satisfied for the real unknown dynamics.

Now, we can formulate a CBF-based QP in the form:
\begin{equation}\label{eqn:prob_qp}
\min_{\bm u(t), \delta(t)} \int_{0}^T ||\bm u(t)||^2 + p\delta^2(t) dt
\end{equation}
subject to (\ref{eqn:cbf_re}), (\ref{eqn:control}), and the CLF constraint
\begin{equation} \label{eqn:clf_oc}
L_{ f_a}V(\bar{\bm{x}})+L_{g_a}V(\bar{\bm{x}})\bm u + \epsilon V(\bar{\bm{x}})\leq \delta(t),
\end{equation}
where $V(\bar{\bm{x}}) = ||\bm \bar{\bm{x}} - \bm K||^2$, $c_3 = \epsilon > 0$ in Def. \ref{def:clf}, $p > 0$, $\delta(t)$ is a relaxation for the CLF constraint. 

Following the approach introduced at the end of Sec. \ref{sec:pre}, we solve the QP (\ref{eqn:prob_qp}) at time $t_k, k = 1,2\dots$. However, at time $t_k$, the QP (\ref{eqn:prob_qp}) does not generally know the error state $\bm e(t)$ and its derivative $\dot{\bm e}(t), \forall t > t_k$. Thus, it cannot guarantee that the CBF constraint (\ref{eqn:cbf_re}) is satisfied in the time interval $(t_k, t_{k+1}]$, where $t_{k+1}$ is the next time instant to solve the QP. In order to find a condition that guarantees the satisfaction of (\ref{eqn:cbf_re}) $\forall t\in (t_k, t_{k+1}]$, we first let $\bm e = (e_1, \dots, e_n)$ and $\dot{\bm e} = (\dot e_1, \dots, \dot e_n)$ be bounded by  $\bm w = (w_1,\dots, w_n)\in \mathbb{R}_{>0}^n$ and $\bm \nu = (\nu_1,\dots, \nu_n)\in \mathbb{R}_{>0}^n$:
\begin{equation} \label{eqn:eb}
\begin{aligned}
| e_i|\leq  w_i, \qquad
|\dot{ e}_i| \leq \nu_i, \qquad i\in\{1,\dots,n\}.
\end{aligned}
\end{equation}
 These two inequalities can be rewritten in the form $|\bm e|\leq \bm w, |\dot{\bm e}| \leq \bm \nu$ for notational simplicity.

We now consider the state $\bar{\bm x}$ at time $t_k$, which satisfies:
\begin{equation} \label{eqn:statevar}
\bar{\bm x}(t_k) - \bm s \leq \bar{\bm x}\leq \bar{\bm x}(t_k) + \bm s,
\end{equation}
where the inequalities are interpreted componentwise and $\bm s\in \mathbb{R}_{>0}^n$. The choice of $\bm s$ will be discussed later. We denote the set of states that satisfy (\ref{eqn:statevar}) at time $t_k$ by 
\begin{equation} \label{eqn:state_set}
S(t_k) = \{\bm y\in X: \bar{\bm x}(t_k) - \bm s \leq \bm y\leq \bar{\bm x}(t_k) + \bm s\}.
\end{equation}

Now, with (\ref{eqn:eb}) and (\ref{eqn:statevar}), we are ready to find a condition that guarantees the satisfaction of (\ref{eqn:cbf_re}) in the time interval $(t_k, t_{k+1}]$. This is done by considering the minimum value of each component in (\ref{eqn:cbf_re}), as shown next.

 In (\ref{eqn:cbf_re}), let  $b_{f_a,min}(t_k)\in \mathbb{R}$ be the minimum value of $\frac{\partial b(\bar{\bm x} + \bm e)}{\partial \bar{\bm x}} {f_a}(\bar{\bm x})$ for the preceding time interval that satisfies $\bm y\in S(t_k), |\bm e|\leq \bm w, \bm y + \bm e \in C_1$ starting at time $t_k$, i.e., let
{\small\begin{equation} \label{eqn:limit_1}
b_{f_a,min}(t_k) = \min_{\bm y\in S(t_k), |\bm e|\leq \bm w, \bm y + \bm e \in C_1} \frac{\partial b(\bm y + \bm e)}{\partial \bm y} {f_a}(\bm y)
\end{equation}}
Similarly, we can also find the minimum value $b_{\alpha_1,min}(t_k)\in \mathbb{R}$ and $b_{e,min}(t_k)\in \mathbb{R}$ of $\alpha_1(b(\bm x))$ and $\frac{\partial b(\bm x)}{\partial\bm e}\dot{\bm e}$, respectively, for the preceding time interval that satisfies $\bm y\in S(t_k), |\bm e|\leq \bm w, \bm y + \bm e \in C_1, |\dot{\bm e}| \leq \bm \nu$ starting at time $t_k$, i.e., let 
{\small\begin{equation}\label{eqn:limit_2}
b_{\alpha_1,min}(t_k) = \min_{\bm y\in S(t_k), |\bm e|\leq \bm w, \bm y + \bm e \in C_1} \alpha_1(b(\bm y + \bm e))
\end{equation}
\begin{equation}\label{eqn:limit_3}
b_{e,min}(t_k) = \min_{\bm y\in S(t_k), |\bm e|\leq \bm w, |\dot{\bm e}| \leq \bm \nu, \bm y + \bm e \in C_1} \frac{\partial b(\bm y + \bm e)}{\partial\bm e}\dot{\bm e}
\end{equation}
}
For the remaining term in (\ref{eqn:cbf_re}), if $\frac{\partial b(\bm x)}{\partial\bar{\bm x}}g_a(\bar{\bm x})$ is independent of $\bar{\bm x}$ and $\bm e$, then we do not need to find its limit value within the bound $\bm y\in S(t_k), |\bm e|\leq \bm w, \bm y + \bm e \in C_1$; otherwise, 
let $\bar{\bm  x} = (\bar x_1, \dots, \bar x_n) \in \mathbb{R}^n$, $\bm u = (u_1, \dots, u_q) \in \mathbb{R}^q$ and ${g_a} = ({g}_1,\dots, {g}_q)\in \mathbb{R}^{n\times q}$. We make the following assumption:
\begin{assump}\label{asp:sign}
	The sign of $u_i(t_{k+1})$ is the same as $u_i(t_k), i \in\{1,\dots, q\}, k = 1,2\dots$.
\end{assump}
 We can then determine the limit value $b_{g_i,lim}(t_k) \in \mathbb{R}, i \in\{1,\dots, q\}$ of $\frac{\partial b(\bm x)}{\partial\bar{\bm x}} g_i(\bar{\bm x})$ by
{\small\begin{equation}\label{eqn:limit_4}
b_{g_i,lim}(t_k) \!=\! \left\{\!\!\!\!\!\begin{array}{c}  
\mathop{\min}\limits_{\bm y\in S(t_k), |\bm e|\!\leq\! \bm w, \bm y \!+\! \bm e \in C_1}\!\!\!\!\! \frac{\partial b(\bm y \!+\! \bm e)}{\partial\bar{\bm x}} g_i(\bar{\bm x}), \text{ if } u_i(t_k) \!\geq\! 0,\\
\mathop{\max}\limits_{\bm y\in S(t_k), |\bm e|\leq \bm w, \bm y + \bm e \in C_1}\!\!\! \frac{\partial b(\bm y \!+\! \bm e)}{\partial\bar{\bm x}} g_i(\bar{\bm x}), \text{ otherwise }
\end{array} \right.
\end{equation}
}Let $b_{g_a,lim}(t_k) = (b_{g_1,lim}(t_k), \dots, b_{g_q,lim}(t_k))\in\mathbb{R}^{ q}$, and we set $b_{g_a,lim}(t_k) = \frac{\partial b(\bm x)}{\partial\bar{\bm x}}g_a(\bar{\bm x})$ if $\frac{\partial b(\bm x)}{\partial\bar{\bm x}} g(\bar{\bm x})$ is independent of $\bar{\bm x}$ and $\bm e$ for notational simplicity. 

The condition that guarantees the satisfaction of (\ref{eqn:cbf_re}) in the time interval $(t_k, t_{k+1}]$ is then given by
{\small\begin{equation} \label{eqn:cbf_discrete}
b_{f_a,min}(t_k) + b_{g_a,lim}(t_k)\bm u(t_k) + b_{e,min}(t_k) + b_{\alpha_1,min}(t_k) \geq 0.
\end{equation}
}
In order to apply the above condition to the QP (\ref{eqn:prob_qp}), we just replace (\ref{eqn:cbf_re}) by (\ref{eqn:cbf_discrete}), i.e., we have
{\small\begin{equation}\label{eqn:prob_qp_event}
\min_{\bm u(t_k), \delta(t_k)} \int_{0}^T ||\bm u(t_k)||^2 + p\delta^2(t_k) dt
\end{equation}
}subject to (\ref{eqn:cbf_discrete}), (\ref{eqn:control}) and (\ref{eqn:clf_oc}).

Based on the above, we define three events that determine the condition that triggers an instance of solving the QP (\ref{eqn:prob_qp_event}):
\begin{itemize}
	\item \textbf{Event 1:} $|\bm e| \leq \bm w$ is about to be violated.
	\item \textbf{Event 2:} $|\dot{\bm e}| \leq \bm \nu$ is about to be violated.
	\item \textbf{Event 3:} the state of (\ref{eqn:affine_nom}) reaches the boundaries of $S(t_k)$.
\end{itemize}

In other words, the next time instant $t_{k+1}, k = 1, 2\dots$ to solve the QP (\ref{eqn:prob_qp_event}) is determined by:
{\small\begin{equation} \label{eqn:next_time}
\begin{aligned}
t_{k+1} = \min\{t>t_k: |\bm e(t)| = \bm w \text{ or } |\dot{\bm e}(t)| = \bm \nu \\\text{ or } |\bar{\bm x}(t) - \bar{\bm x}(t_k)| = \bm s\},
\end{aligned}
\end{equation}
}where $t_1 = 0$. The first two events can be detected by direct sensor measurements, while Event 3 can be detected by monitoring the dynamics (\ref{eqn:affine_nom}). The magnitude of each component of $\bm s$ is a tradeoff between the time complexity and the conservativeness of this approach. If the magnitude is large, then the number of events is small but this approach is considerably conservative as we determine the condition (\ref{eqn:cbf_discrete}) through the minimum values as in (\ref{eqn:limit_1})-(\ref{eqn:limit_4}).

Formally, we have the following theorem to show that the satisfaction of the safety constraint (\ref{eqn:safetycons}) is guaranteed for the real unknown dynamics with the condition (\ref{eqn:cbf_discrete}):
\begin{theorem} \label{thm:cbf_discrete}
	Given a HOCBF $b(\bm x)$ with $m = 1$ as in Def. \ref{def:hocbf}, let $t_{k+1}, k = 1, 2\dots$ be determined by (\ref{eqn:next_time}) with $t_1 = 0$, and $b_{f_a,min}(t_k), b_{\alpha_1,min}(t_k), b_{e,min}(t_k), b_{g_a,lim}(t_k)$ be determined by (\ref{eqn:limit_1})-(\ref{eqn:limit_4}), respectively. Then, under Assumption \ref{asp:sign}, any control $\bm u(t_k)$ that satisfies
	(\ref{eqn:cbf_discrete})
	and updates the real unknown dynamics and the adaptive dynamics (\ref{eqn:affine_nom}) within time interval $[t_k, t_{k+1})$ renders the set $C_1$ forward invariant for the real unknown dynamics.
\end{theorem}
\textbf{Proof:} By (\ref{eqn:next_time}), we have that 
$$
\bm y(t)\in S(t_k), |\bm e(t)|\leq \bm w, |\dot{\bm e}(t)| \leq \bm \nu, \bm y(t) + \bm e(t) \in C_1
$$
for all $t\in[t_k, t_{k+1}], k = 1, 2\dots$. Thus, the limit values $b_{f_a,min}(t_k), b_{\alpha_1,min}(t_k), b_{e,min}(t_k)$, determined by (\ref{eqn:limit_1}), (\ref{eqn:limit_2}), (\ref{eqn:limit_3}), respectively, are the minimum values for $\frac{\partial b(\bm x)}{\partial\bar{\bm x}}f_a(\bar{\bm x}(t)), \alpha_1(b(\bm x(t))), \frac{\partial b(\bm x(t))}{\partial\bm e(t)} \dot{\bm e}(t)$ for all $t\in[t_k, t_{k+1})$. In other words, we have
{\small$$
\begin{aligned}
\frac{\partial b(\bm x)}{\partial\bar{\bm x}}f_a(\bar{\bm x}(t)) + \alpha_1(b(\bm x(t)))+\frac{\partial b(\bm x(t))}{\partial\bm e(t)} \dot{\bm e}(t)\geq  b_{f_a,min}(t_k)\\+ b_{\alpha_1,min}(t_k)+ b_{e,min}(t_k), \forall t\in[t_k,t_{k+1}).
\end{aligned}
$$
}where $\bm x = \bm e + \bar{\bm x}$.

By Assumption \ref{asp:sign}, we have that
{\small$$
\frac{\partial b(\bm x)}{\partial\bar{\bm x}}g_a(\bar{\bm x}(t))\bm u(t_k) \geq b_{g_a,lim}(t_k)\bm u(t_k), \forall t\in[t_k,t_{k+1}).
$$
}
Following the last two equations and (\ref{eqn:cbf_discrete}), we have
{\small$$
\begin{aligned}
	\frac{\partial b(\bm x)}{\partial\bar{\bm x}} f_a(\bar{\bm x}(t)) + \frac{\partial b(\bm x)}{\partial\bar{\bm x}} g_a(\bar{\bm x}(t))\bm u(t_k)+\alpha_1(b(\bm x(t)))\\+\frac{\partial b(\bm x(t))}{\partial\bm e(t)} \dot{\bm e}(t)\geq  0, \forall t\in[t_k,t_{k+1}).
\end{aligned}
$$
}
By (\ref{eqn:cbf}), (\ref{eqn:cbf_re}) and the last equation, we have that 
{\small$$
\begin{aligned}
\frac{db(\bar{\bm x}(t) + \bm e(t))}{dt} + \alpha_1(b(\bar{\bm x}(t) + \bm e(t)))\geq 0, \\ \forall t\in[t_k,t_{k+1}), k = 1,2\dots
\end{aligned}
$$
}
By Thm. \ref{thm:hocbf}, we have that $b(\bar{\bm x}(t) + \bm e(t)) \geq 0, \forall t\geq 0$ and by (\ref{eqn:alt}), we have that $C_1$ is forward invariant for the real unknown dynamics. $\qquad\qquad\qquad\qquad\qquad\qquad\qquad\qquad\;\blacksquare$

Note that if Assumption \ref{asp:sign} is not satisfied, we can consider the alternative case in (\ref{eqn:limit_4}). 

\begin{remark}\label{rem:break}
	We may also consider the minimum value of $\frac{\partial b(\bm y + \bm e)}{\partial \bm y} f_a(\bm y) + \frac{\partial b(\bm y + \bm e)}{\partial\bm e} \dot{\bm e} + \alpha_1(b(\bm y + \bm e))$ within the bounds $\bm y\in S(t_k), |\bm e|\leq \bm w, \bm y + \bm e \in C_1, |\dot{\bm e}| \leq \bm \nu$ instead of considering them separately as in (\ref{eqn:limit_1})-(\ref{eqn:limit_4}). However, this could be too conservative as the constraint (\ref{eqn:cbf_discrete}) is stronger compared with the CBF constraint (\ref{eqn:cbf_re}), and we wish to find the largest possible value of the left-hand side of (\ref{eqn:cbf_re}) that can support the proof of Thm. \ref{thm:cbf_discrete}.  To reduce conservativeness, we may further break (\ref{eqn:limit_1})-(\ref{eqn:limit_4}) into smaller components (an example can be found in Sec. \ref{sec:case}) and find the corresponding minimum values that are then included in a conditon similar to (\ref{eqn:cbf_discrete}).
\end{remark}

Events 1 and 2 will be frequently triggered if the modelling of the adaptive dynamics (\ref{eqn:affine_nom}) has a large error with respect to the real dynamics. Therefore, we would like to model the adaptive dynamics (\ref{eqn:affine_nom}) as accuracy as possible in order to reduce the number of events required to solve the  QP (\ref{eqn:prob_qp_event}). 

An additional important step is to synchronize the state of the real unknown dynamics and (\ref{eqn:affine_nom}) such that we always have $\bm e(t_k) = 0$ and make $\dot{\bm e}(t_k)$ close to 0 by setting 
\begin{equation} \label{eqn:state_reset}
\bar{\bm x}(t_k) = \bm x(t_k),
\end{equation}
and by updating $f_a(\bar{\bm{x}}(t))$ of the adaptive dynamics (\ref{eqn:affine_nom}) right after ($t^+$) an event occurs at $t$:
{\small\begin{equation}
f_a(\bar{\bm{x}}(t^+))= f_a(\bar{\bm{x}}(t^-)) + \sum_{i = 0}^k\dot{\bm e}(t_i). \label{eqn:affine_nom_up}%
\end{equation}
}where $t^+, t^-$ denote instants right after and before $t$. In this way, the dynamics (\ref{eqn:affine_nom}) are adaptively updated at each event, i.e., at $t_k, k = 1, 2,\dots$. Note that we may also update $g_a(\cdot)$, which is harder than updating $f_a(\cdot)$  since $g_a(\cdot)$ is multiplied by $\bm u$ that is to be determined, i.e., the update of $g_a(\cdot)$ will depend on $\bm u$. This possibility is the subject of ongoing work.

By (\ref{eqn:state_reset}) and (\ref{eqn:affine_nom_up}), 
we have that $\bm e(t_k) = 0$ and $\dot{\bm e}(t_k)$ is close to 0. There exist lower bounds for the occurance times of Event 1 and Event 3, and they are determined by the limit values of the component of $f_a, g_a$ within $X$ and $U$, as well as the real unknown dynamics (although they are unknown). Assuming the functions that define the real unknown dynamics are Lipschitz continuous, and the functions $f_a, g_a$ in (\ref{eqn:affine_nom}) are also assumed to be Lipschitz continuous, it follows that $\dot{\bm e}$ is also Lipschitz continuous. Suppose the largest Lipschitz constant among all the components in $\dot{\bm x}$ is $L, \forall \bm x\in X$, and the smallest Lipschitz constant among all the components in $\dot{\bar{\bm x}}$ is $\bar L, \forall \bar{\bm x}\in X$, then the lower bound time for the Event 2 is $\frac{\nu_{min}}{L - \bar L}$, where $\nu_{min} > 0$ is the minimum component in $ \bm \nu$.
We summarize the event-triggered control in Alg. \ref{alg:et}.

\begin{algorithm}
		\caption{Event-triggered control} \label{alg:et}
		\KwIn{Measurements $\bm x$ and $\dot{\bm x}$ from a plant, adaptive model (\ref{eqn:affine_nom}), settings for QP (\ref{eqn:prob_qp_event}), $\bm w, \bm \nu, \bm s$.}
		\KwOut{Event time $t_k, k = 1, 2, \dots$ and $\bm u^*(t_k)$.}
		$k = 1, t_k = 0$\;
		\While{$t_k \leq T$}{
		Measure $\bm x$ and $\dot{\bm x}$ from the plant at $t_k$\;
		Sync. the state of (\ref{eqn:affine_nom}) and the plant by (\ref{eqn:state_reset}),(\ref{eqn:affine_nom_up})\;
		Evaluate (\ref{eqn:limit_1})-(\ref{eqn:limit_4})\;
		Solve the QP (\ref{eqn:prob_qp_event}) at $t_k$ and get $\bm u^*(t_k)$\;	
		\While{$t\leq T$}{			
			Apply $\bm u^*(t_k)$ to the plant and (\ref{eqn:affine_nom}) for $t\geq t_k$\;	
			Measure $\bm x$ and $\dot{\bm x}$ from the plant\;
			Evaluate $t_{k+1}$ by (\ref{eqn:next_time})\;		
			\If{ $t_{k+1}$ is found with $\varepsilon > 0$ error}{
				$k\leftarrow k+1$, break\;
			}
		}
	}
\end{algorithm}

\begin{remark}\label{rem:uncer}(Measurement uncertainties)
	If the measurements $\bm x$ and $\dot{\bm x}$ are subject to uncertainties, and the uncertainties are bounded, then we can apply the bounds of $\bm x$ and $\dot{\bm x}$ in evaluating $t_{k+1}$ by (\ref{eqn:next_time}) instead of $\bm x$ and $\dot{\bm x}$ themselves. In other words, $\bm e(t)$ and $\dot{\bm e}(t)$ are determined by the bounds of $\bm x$, $\dot{\bm x}$ and the state values of the adaptive system (\ref{eqn:affine_nom}).
\end{remark}

\subsection{High-relative-degree Constraints}

In this subsection, we consider the safety constraint (\ref{eqn:safetycons}) whose relative degree is larger than one with respect to the real unknown dynamics and (\ref{eqn:affine_nom}). In other words, we need to consider the HOCBF constraint (\ref{eqn:constraint}) to find the state feedback control with the HOCBF method.

Similar to the last subsection, we find the error state $\bm e$ by (\ref{eqn:err}), and have an alternative form of the HOCBF $b(\bm x)$ as in (\ref{eqn:alt}). The HOCBF constraint (\ref{eqn:constraint}) that guarantees $b(\bar{\bm x} + \bm e)\geq 0$ with respect to the real unknown dynamics is
{\small\begin{equation}\label{eqn:hocbf_alt}
\begin{aligned} 
\frac{\partial^m b(\bm x)}{\partial\bar{\bm x}^m} f_a(\bar{\bm x}) + \frac{\partial^m b(\bm x)}{\partial\bar{\bm x}^m} g_a(\bar{\bm x})\bm u + \frac{\partial^m b(\bm x)}{\partial {\bm e}^m}{\bm e}^{(m)} \\+ R(b(\bm x)) + \alpha_m(\psi_{m-1}(\bm x))
 \geq 0, \end{aligned}
\end{equation}
}where $R(b(\bm x))$ also contains the remaining time derivatives of $\bm e$ with degree less than $m$. ${\bm e}^{(i)} = {\bm x}^{(i)} - {\bar{\bm x}}^{(i)}, i\in\{1,\dots,m\}$ is the $i_{th}$ derivative and is evaluated online by ${\bm x}^{(i)}$ (from a sensor) of the real system and $\bar{\bm x}^{(i)}$ of (\ref{eqn:affine_nom}).

In order to find a conditon that guarantees the satisfaction of the last equation in $[t_i,t_{i+1}),i = 1, 2,\dots$, we let $\bm e$ and ${\bm e}^{(i)}, i\in\{1,\dots,m\}$ be bounded by  $\bm w\in \mathbb{R}_{>0}^n$ and $\bm \nu_i\in \mathbb{R}_{>0}^n$, i.e., we have
\begin{equation} \label{eqn:err_bnd}
\begin{aligned}
|\bm e|&\leq \bm w,\qquad
|{\bm e}^{(i)}| &\leq \bm \nu_i, i\in\{1,\dots,m\},
\end{aligned}
\end{equation}
where the inequalities are interpreted componentwise and the absolute function $|\cdot|$ applies to each component.  We also consider the set of states in the form of (\ref{eqn:state_set}).

We further define a set $S_h(t_k)$in the form
{\small\begin{equation}
\begin{aligned}
S_h(t_k) = \{\bm y, \bm e, {\bm e}^{(1)}, \dots, {\bm e}^{(m)}: \bm y\in S(t_k), |\bm e|\leq \bm w,\\ \wedge_{i=1}^m(|{\bm e}^{(i)}| \leq \bm \nu_i), \bm y + \bm e \in \cap_{i=1}^m C_i\}
\end{aligned}
\end{equation}
}where $\wedge_{i=1}^m$ denotes the conjunction from $1$ to $m$.

Then, we can find the minimum values $b_{f_a^m,min}(t_k)\in \mathbb{R}, b_{\alpha_m,min}(t_k)\in \mathbb{R}, b_{e^m,min}(t_k)\in \mathbb{R}, b_{R,min}(t_k)\in \mathbb{R}$ for the preceding time interval that satisfies $(\bm y, \bm e, {\bm e}^{(1)}, \dots, {\bm e}^{(m)}) \in S_h(t_k)$ starting at time $t_k$ by 
{\small\begin{equation} \label{eqn:limit_H1}
b_{f_a^m,min}(t_k) = \min_{(\bm y, \bm e, {\bm e}^{(1)}, \dots, {\bm e}^{(m)}) \in S_h(t_k)} \frac{\partial^m b(\bm y + \bm e)}{\partial {\bm y}^m} {f_a}(\bm y)
\end{equation}
\begin{equation}\label{eqn:limit_H2}
b_{\alpha_m,min}(t_k) = \min_{(\bm y, \bm e, {\bm e}^{(1)}, \dots, {\bm e}^{(m)}) \in S_h(t_k)} \alpha_m(\psi_{m-1}(\bm y + \bm e))
\end{equation}
\begin{equation}\label{eqn:limit_H3}
b_{e^m,min}(t_k) = \min_{(\bm y, \bm e, {\bm e}^{(1)}, \dots, {\bm e}^{(m)}) \in S_h(t_k)} \frac{\partial^m b(\bm y + \bm e)}{\partial{\bm e}^m}{\bm e}^{(m)}
\end{equation} 
\begin{equation}\label{eqn:limit_H31}
b_{R,min}(t_k) = \min_{(\bm y, \bm e, {\bm e}^{(1)}, \dots, {\bm e}^{(m)}) \in S_h(t_k)} R(b(\bm y + \bm e))
\end{equation}
}
If $\frac{\partial^m b(\bm x)}{\partial{\bar{\bm x}}^m} g_a(\bar{\bm x})$ is independent of $\bar {\bm x}$ and $\bm e$, then we do not need to find its limit value within the set $S_h(t_k)$; otherwise, we can determine its limit value $b_{g_i,lim}(t_k) \in \mathbb{R}, i \in\{1,\dots, q\}$ under Assumption \ref{asp:sign} by
{\small\begin{equation}\label{eqn:limit_H4}
b_{g_i,lim}(t_k) \!= \!\!\!\left\{\!\!\!\!\begin{array}{c}  
\mathop{\min}\limits_{(\bm y, \bm e, {\bm e}^{(1)} \dots {\bm e}^{(m)}) \in S_h(t_k)}\!\!\!\!\!\! \frac{\partial^m b(\bm y + \bm e)}{\partial{\bm y}^m} g_i(\bm y), \text{if }\! u_i(t_k) \!\geq\! 0,\\
\mathop{\max}\limits_{(\bm y, \bm e, {\bm e}^{(1)} \dots {\bm e}^{(m)}) \in S_h(t_k)}\!\!\!\!\!\! \frac{\partial^m b(\bm y + \bm e)}{\partial{\bm y}^m} g_i(\bm y), \text{ otherwise }
\end{array} \right.
\end{equation}
}Let $b_{g_a,lim}(t_k) = (b_{g_1,lim}(t_k), \dots, b_{g_q,lim}(t_k))\in\mathbb{R}^{1\times q}$, and we set $b_{g_a,lim}(t_k) = \frac{\partial^m b(\bm x)}{\partial{\bar{\bm x}}^m} g_a(\bar{\bm x})$ if $\frac{\partial^m b(\bm x)}{\partial{\bar{\bm x}}^m} g_a(\bar{\bm x})$ is independent of $\bar {\bm x}$ and $\bm e$ for notational simplicity.

Similar to Remark \ref{rem:break}, we can break the above terms into smaller components (an example can be found in the case study section) and finding their corresponding minimum values in order to make this approach less conservative.  The condition that guarantees the satisfaction of (\ref{eqn:hocbf_alt}) in the time interval $[t_k, t_{k+1})$ is then given by
{\small\begin{equation} \label{eqn:hocbf_discrete}
\begin{aligned}
b_{f_a^m,min}(t_k) + b_{g_a,lim}(t_k)\bm u(t_k) + b_{e^m,min}(t_k) \\+ b_{\alpha_m,min}(t_k) + b_{R,min}(t_k) \geq 0.
\end{aligned}
\end{equation}
}
In order to apply the above condition to the QP (\ref{eqn:prob_qp}), we just replace (\ref{eqn:cbf_re}) by (\ref{eqn:hocbf_discrete}), i.e., we have
{\small\begin{equation}\label{eqn:prob_qp_Hevent}
\min_{\bm u(t_k), \delta(t_k)} \int_{0}^T ||\bm u(t_k)||^2 + p\delta^2(t_k) dt
\end{equation}
}subject to (\ref{eqn:hocbf_discrete}), (\ref{eqn:control}) and (\ref{eqn:clf_oc}).

 Based on the above, we also have three events that determine the trigger of solving the QP (\ref{eqn:prob_qp_Hevent}):
\begin{itemize}
	\item \textbf{Event 1:} $|\bm e| \leq \bm w$ is about to be violated.
	\item  \textbf{Event 2:} $|{\bm e}^{(i)}| \leq \bm \nu_i$ is about to be violated for each $i\in\{1,\dots,m\}$.
	\item \textbf{Event 3:} the state of (\ref{eqn:affine_nom}) reaches the boundaries of $S(t_k)$.
\end{itemize}

The next time instant $t_{k+1}, k = 1, 2\dots$ ($t_1 = 0$) to solve the QP (\ref{eqn:prob_qp_Hevent}) is determined by:
{\small\begin{equation} \label{eqn:next_timeH}
\begin{aligned}
t_{k+1} = \min\{t>t_k: |\bm e(t)| = \bm w \text{ or } |{\bm e}^{(i)}(t)| = \bm \nu_i,\\ i\in\{1,\dots,m\}\text{ or } |\bar{\bm x}(t) - \bar{\bm x}(t_k)| = \bm s\},
\end{aligned}
\end{equation}
}
Formally, we have the following theorem that shows the satisfaction of the safety constraint (\ref{eqn:safetycons}) for the real unknown dynamics:
\begin{theorem} \label{thm:hocbf_discrete}
	Given a HOCBF $b(\bm x)$ as in Def. \ref{def:hocbf}. Let $t_{k+1}, k = 1, 2\dots$ be determined by (\ref{eqn:next_timeH}) with $t_1 = 0$, and $b_{f_a^m, min}(t_k), b_{\alpha_m, min}(t_k), b_{e^m, min}(t_k), b_{R, min}^m(t_k), b_{g_a, lim}(t_k)$ be determined by (\ref{eqn:limit_H1})-(\ref{eqn:limit_H4}), respectively. Then, under Assumption \ref{asp:sign}, any control $\bm u(t_k)$ that satisfies
	(\ref{eqn:hocbf_discrete})
	and updates the real unknown dynamics and (\ref{eqn:affine_nom}) within time interval $[t_k, t_{k+1})$ renders the set $C_1\cap\dots\cap C_m$ forward invariant for the real unknown dynamics.
\end{theorem}
\textbf{Proof:} Similar to the proof of Thm. \ref{thm:cbf_discrete}, we have 
{\small$$
\begin{aligned}
 \frac{\partial^m b(\bm x(t))}{\partial t^m} + R(b(\bm x(t))) + \alpha_m(\psi_{m-1}(\bm x(t))) \geq 0,\\ \forall t\in[t_k,t_{k+1}], k = 1,2\dots
\end{aligned}
$$
}which is equivalent to the HOCBF constraint (\ref{eqn:constraint}) in Def. \ref{def:hocbf}. Then, by Thm. \ref{thm:hocbf}, (\ref{eqn:alt}) and ${\bm e}^{(i)} = {\bm x}^{(i)} - {\bar{\bm x}}^{(i)}, i\in\{1,\dots,m\}$, we can recursively show that $C_i, i\in\{1,\dots, m\}$ are forward invariant, i.e., the set $C_1\cap\dots\cap C_m$ is forward invariant for the real unknown dynamics. $\qquad\qquad\qquad\qquad\qquad\qquad\blacksquare$

If Assumption \ref{asp:sign} is not satisfied, we can also consider the alternative case in (\ref{eqn:limit_H4}). 
The process can be summarized through an algorithm similar to Alg. \ref{alg:et} and we can deal with measurement uncertainties as in Remark \ref{rem:uncer}. We can also synchronize the state of the real unknown dynamics and (\ref{eqn:affine_nom}) as in (\ref{eqn:state_reset}) and (\ref{eqn:affine_nom_up}) such that we always have $\bm e(t_k) = 0$ and ${\bm e}^{(i)}(t_k), i\in \{1,\dots,m\}$ stay close to 0.
\section{CASE STUDIES}

\label{sec:case}

In this section, we consider the case study of an ACC problem. All the computations and simulations were conducted in MATLAB. We used quadprog to solve the quadratic programs and ode45 to integrate the dynamics.

The real vehicle dynamics are \textbf{unknown} to the controller:
{\small\begin{equation}
\left[
	\begin{array}
	[c]{c}%
	\dot{v}(t)\\
	\dot z(t)
	\end{array}
	\right]  =\left[
	\begin{array}
	[c]{c}%
	\sigma_1(t) + \frac{\sigma_3(t)}{M}u(t) -\frac{1}{M}F_{r}(v(t))\\
	\sigma_2(t) + v_p - v(t)
	\end{array}
	\right]   \label{eqn:vehicle}%
\end{equation}
}where $\bm x = (v, z)$ and $z(t)$ denotes the distance between the preceding and the ego vehicle, $v_p > 0, v(t)$ denote the velocities of the  preceding and ego
vehicles along the lane (the velocity of the preceding vehicle is assumed constant), respectively, and $u(t)$ is the control of the ego vehicle. $\sigma_1(t), \sigma_2(t),\sigma_3(t)$ denote three random processes whose pdf's have finite support. $M$ denotes the mass of the ego vehicle and $F_{r}(v(t))$
denotes the resistance force, which is expressed \cite{Khalil2002} as:
$F_{r}(v(t))=f_{0}sgn(v(t))+f_{1}v(t)+f_{2}v^{2}(t),$
where $f_{0} > 0,$
$f_{1} > 0$ and $f_{2}> 0$ are unknown. 

The adaptive dynamics will be automatically updated as shown in (\ref{eqn:affine_nom_up}), and are in the form:
{\small\begin{equation}
\underbrace{\left[
	\begin{array}
	[c]{c}%
	\dot{\bar v}(t)\\
	\dot {\bar z}(t)
	\end{array}
	\right]  }_{\dot{\bar{\bm x}}(t)}=\underbrace{\left[
	\begin{array}
	[c]{c}%
	h_1(t)-\frac{1}{M}F_{n}(\bar v(t))\\
	h_2(t) + v_p - \bar v(t)
	\end{array}
	\right]  }_{f_a(\bar{\bm x}(t))}+\underbrace{\left[
	\begin{array}
	[c]{c}%
	\frac{1}{M}\\
	0
	\end{array}
	\right]  }_{g_a(\bar{\bm x}(t))}u(t)\label{eqn:vehicle_norm}%
\end{equation}
}where $h_1(t)\in\mathbb{R}, h_2(t)\in \mathbb{R}$ denote the two adaptive terms in (\ref{eqn:affine_nom_up}) (also see (\ref{eqn:veh_nom_up})), $h_1(0) = 0, h_2(0) = 0$. $\bar z(t)$ denotes the distance between the preceding and the ego vehicle for the above adaptive dynamics, and $\bar v(t)$ denotes the velocity. $F_{n}(\bar v(t))$
denotes the resistance force, which is \textit{different} from $F_r$ in (\ref{eqn:vehicle}) and is expressed as:
$F_{n}(\bar v(t))=g_{0}sgn(\bar v(t))+g_{1}\bar v(t)+g_{2}\bar v^{2}(t),$ 
where $g_{0}>0, $
$g_{1} > 0$ and $g_{2}>0$ are empirically determined.

The control bound is defined as:
$
-c_d Mg\leq u(t)\leq c_a Mg,\quad\forall t\geq0, 
$
where $c_a > 0$ and $c_d>0$ are the maximum acceleration and deceleration coefficients, respectively.

We require that the distance $z(t)$ between the ego vehicle (real dynamics) and its immediately
preceding vehicle be greater than $l_{p} > 0$,
i.e.,
\begin{equation}
\label{eqn:safety}z(t) \geq l_{p},\quad\forall t\geq0.
\end{equation}

The objective is to minimize the cost $\int_{0}^{T}\left(\frac{u(t)-F_{r}(v(t))}{M}\right)^{2}dt$. The ego vehicle is also trying to achieve a desired speed $v_d > 0$, which is implemented by a CLF $V(\bar{\bm x}) = (\bar v - v_d)^2$ with $c_3 = \epsilon$ as in Def. \ref{def:clf}. Since the relative degree of the rear-end safety constraint (\ref{eqn:safety}) is two, we define a HOCBF $b(\bm x) = z - l_p$ with $\alpha_1(b(\bm x)) = b(\bm x)$ and $\alpha_2(\psi_1(\bm x)) = \psi_1(\bm x)$ as in Def. \ref{def:hocbf} to implement the safety constraint. Then, the HOCBF constraint (\ref{eqn:constraint}) which in this case is (with respect to the real dynamics (\ref{eqn:vehicle})):
$\ddot b(\bm x) + 2\dot b(\bm x) + b(\bm x)\geq 0.$
Combining (\ref{eqn:err}), (\ref{eqn:vehicle_norm}) and this equation, we have a HOCBF constraint in the form:
{\small\begin{equation}
\begin{aligned}
\underbrace{-h_1(t) + \frac{F_n(\bar v(t))}{M}}_{\frac{\partial^2 b(\bm x)}{\partial{\bar{\bm x}^2}} f_a(\bar{\bm x})} + \underbrace{\frac{-1}{M}}_{\frac{\partial^2 b(\bm x)}{\partial{\bar{\bm x}^2}} g_a(\bar{\bm x})} u(t) + \underbrace{\ddot e_2(t)}_{\frac{\partial^2 b(\bm x)}{\partial {\bm e}^2} \ddot {\bm e}} \\+ \underbrace{2(h_2(t) + v_p - \bar v(t) + \dot e_2(t)) + \bar z(t) + e_2(t) - l_p}_{R(b(\bm x)) + \alpha_2(\psi_1(\bm x))} \geq 0
\end{aligned}
\end{equation}
}where $\bm e = (e_1, e_2), e_1 = v - \bar v, e_2 = z - \bar z$.  As in (\ref{eqn:statevar}) and (\ref{eqn:err_bnd}), we consider the state and bound the errors at step $t_k, k = 1,2\dots$ for the above HOCBF constraint in the form:
{\small\begin{equation}
\begin{aligned}
\bar v(t_k) - s_1&\leq \bar v\leq \bar v(t_k) + s_1,\\
\bar z(t_k) - s_2&\leq \bar z\leq \bar z(t_k) + s_2,\\
|e_2|\leq w_2, \quad
|\dot e_2|& \leq \nu_{2,1}, \quad
|\ddot e_2|\leq \nu_{2,2}
\end{aligned}
\end{equation}
}where $s_1 > 0, s_2 > 0, w_2 > 0, \nu_{2,1} > 0, \nu_{2,2}> 0$.

Motivated by (\ref{eqn:state_set}) and (\ref{eqn:affine_nom_up}), we also synchronize the state and update the adaptive dynamics (\ref{eqn:vehicle_norm}) at step $t_k, k = 1,2\dots$ in the form:
{\small
\begin{equation} \label{eqn:veh_nom_up}
\begin{aligned}
\bar v(t_k) = v(t_k),\quad&
\bar z(t_k) = z(t_k),\\
h_1(t^+) \!=\! h_1(t^-) \!-\! \sum_{i = 0}^k\ddot e_2(t_i),&
h_2(t^+)\!=\! h_2(t^-) \!+\! \sum_{i = 0}^k\dot e_2(t_i),
\end{aligned}
\end{equation}}where $\dot e_2(t_k) = \dot z(t_k) - (h_2 + v_p - \bar v (t_k)), \ddot e_2(t_k) = \ddot z(t_k) - \frac{F_n(\bar v(t_k)) - u(t_k^-)}{M} + h_1(t_k), u(t_k^-) = u(t_{k-1})$ and $u(t_0) = 0$. $\dot z(t_k), \ddot z(t_k)$ are estimated by a sensor that measures the ego real dynamics (\ref{eqn:vehicle}) at time $t_k$.

Then, we can find the limit values as in (\ref{eqn:limit_H1})-(\ref{eqn:limit_H4}), solve the QP (\ref{eqn:prob_qp_Hevent}) at each time step $t_k, k = 1,2\dots$, and evaluate the next time step $t_{k+1}$ by (\ref{eqn:next_timeH}) afterwards. In the evaluation of $t_{k+1}$, we have $e_2 = z - \bar z, \dot e_2 = \dot z - (h_2 + v_p - \bar v), \ddot e_2 = \ddot z - \frac{F_n(\bar v) - u(t_k)}{M}  + h_1$, where $z, \dot z, \ddot z$ are estimated by a sensor that measures the ego real dynamics (\ref{eqn:vehicle}), and $u(t_k)$ is already obtained by solving the QP (\ref{eqn:prob_qp_Hevent}) and is held as a constant until we find $t_{k+1}$. The optimization of (\ref{eqn:limit_H1}) is also a QP, while the optimizations of (\ref{eqn:limit_H2})- (\ref{eqn:limit_H4}) are LPs. Therefore, they can all be efficiently solved. Each QP or LP can be solved with a computational time $<0.01s$ in MATLAB (Intel(R) Core(TM) i7-8700 CPU @
3.2GHz$\times2$).

In the simulation, the initial states of the real dynamics (\ref{eqn:vehicle}) and the adaptive dynamics (\ref{eqn:vehicle_norm}) are $\bm x(0) = \bar{\bm x}(0) = (20 m/s, 100 m)$. The final time is $T = 30s$. Other simulation parameters are $v_p = 13.89m/s, v_d = 24m/s, M = 1650kg, g = 9.81m/s^2, f_0 = 0.1N, f_1 = 5Ns/m, f_2 = 0.25Ns^2/m, g_0 = 0.3N, g_1 = 10Ns/m, g_2 = 0.5Ns^2/m, s_1 = 0.4m/s, s_2 = 0.5m, w_2 = 1m, \nu_{2,1} = 0.5m/s, \nu_{2,2} = 0.2m/s^2, c_a = 0.6, c_d = 0.6, p = 1, \epsilon = 10$.

%
	

The pdf's of $\sigma_1(t), \sigma_2(t), \sigma_3(t)$ are uniform over the intervals $[-0.2, 0.2]m/s^2, [-2, 2] m/s, [0.9, 1]$, respectively. The sensor sampling rate is 20Hz. We compare the proposed event driven framework with the time driven approach. The discretization time for the time driven approach is $\Delta t = 0.1$.

\begin{figure}[thpb]
	\centering
	\includegraphics[scale=0.55]{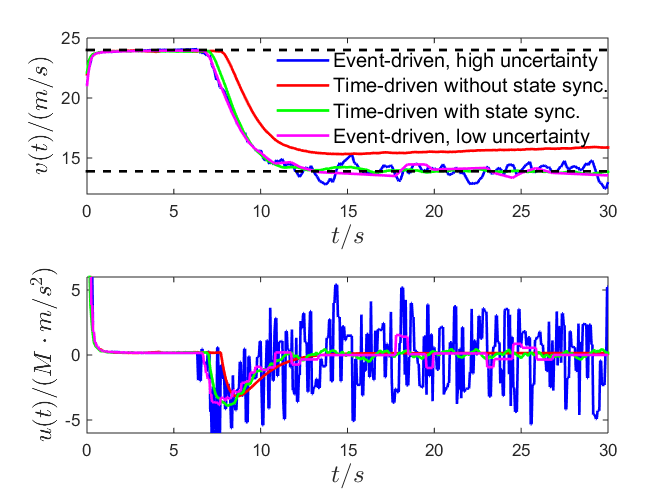}
	\vspace{-3mm}
	\caption{Speed and control profiles for the proposed event driven framework and time driven with or without state synchronization.}
	\label{fig:spd_ctrl}	
\end{figure}

\begin{figure}[thpb]
	\vspace{-3mm}
	\centering
	\includegraphics[scale=0.55]{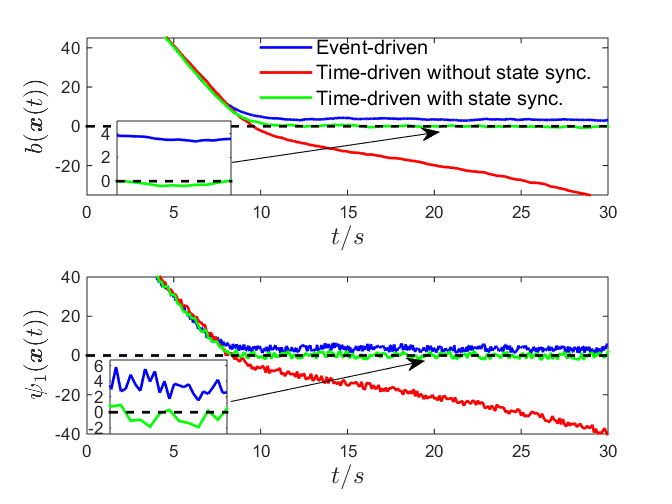}
	\vspace{-3mm}
	\caption{The variations of functions $b(\bm x(t))$ and $\psi_1(\bm x(t))$ for the proposed event driven framework and time driven with or without state synchronization. $b(\bm x(t))\geq 0$ and $\psi_1(\bm x(t))\geq 0$ imply the forward invariance of $C_1\cap C_2$. The set $C_1\cap C_2$ is forward invariant for the real dynamics (\ref{eqn:vehicle}) with the proposed event driven framework. Both the set $C_1$ and $C_2$ are not forward invariant under time driven approach with or without state synchronization. }	
	\label{fig:safety}
\end{figure}

The simulation results are shown in Figs. \ref{fig:spd_ctrl} and \ref{fig:safety}. Note that in the event-driven approach (blue lines), the control varies largely in order to be responsive to the random processes in the real dynamics.  The control is constant in each time interval, and thus the Lipschitz condition is satisfied. If we decrease the uncertainty levels by 10 times, the control is smoother (magenta lines). Thus, highly accurately modelled adaptive dynamics can smooth the control. 

It follows from Fig. \ref{fig:safety}  that the set $C_1\cap C_2$ is forward invariant for the real vehicle dynamics (\ref{eqn:vehicle}), i.e., the safety constraint (\ref{eqn:safety}) is guaranteed with the proposed event driven approach. However, the safety is not guaranteed even with state synchronization under the time-driven approach.

In the event-driven approach, the number of QPs (events) within time $[0, T]$ is reduced by about 50\% compared with the time-driven approach. If we multiply the bounds of the random processes $\sigma_1(t), \sigma_2(t)$ by 2, then the number of events increases by about 23\% for both the 20Hz and 100Hz sensor sampling rate, which shows that accurate adaptive dynamics can reduce the number of events, and thus improves the computational efficient. 

\section{CONCLUSION \& FUTURE WORK}

\label{sec:conclusion}

This paper proposes an event-triggered framework for safety-critical control of systems with unknown dynamics. This framework is based on defining adaptive affine dynamics to estimate the real system, an event-trigger mechanism for solving the problem and the finding of a condition that guarantees safety between events. We have demonstrated the effectiveness of the proposed framework by applying it
to an adaptive cruise control problem. In the future, we will study the issue of guaranteeing the feasibility of all QPs in the proposed framework.






\bibliographystyle{IEEEtran}
\bibliography{CBF}

\end{document}